\documentclass[prd,aps,twocolumn,amsmath,amssymb,nofootinbib,preprintnumbers]
{revtex4}

\usepackage{graphicx}
\usepackage{dcolumn}
\usepackage{bm}
\usepackage{amsmath}
\usepackage{amsthm}

\usepackage{amsfonts}
\usepackage{euscript,bbm}
\usepackage{ifthen}
\usepackage{psfrag}
\usepackage{slashed}

\def\ls{\mathrel{\lower4pt\vbox{\lineskip=0pt\baselineskip=0pt
           \hbox{$<$}\hbox{$\sim$}}}}
\def\gs{\mathrel{\lower4pt\vbox{\lineskip=0pt\baselineskip=0pt
           \hbox{$>$}\hbox{$\sim$}}}}
\def\drawbox#1#2{\hrule height#2pt

\hbox{\vrule width#2pt height#1pt \kern#1pt
              \vrule width#2pt}
              \hrule height#2pt}

\def\Asym#1#2{\vcenter{\vbox{\drawbox{#1}{#2}
              \kern-#2pt       
              \drawbox{#1}{#2}}}}

\newcommand{\be}{\begin{equation}}
\newcommand{\ee}{\end{equation}}
\newcommand{\bea}{\begin{eqnarray}}
\newcommand{\eea}{\end{eqnarray}}

\newcommand{\met} {{E\!\!\!\!/_{\rm T}}}
\newcommand{ \alpgen } {{\tt ALPGEN}}
\newcommand{ \pythia } {{\tt PYTHIA}}

\newcommand{ \isajet }    {{\tt ISAJET}}
\newcommand{ \pgs }    {{\tt PGS4}}

\begin{document}

%
\title{Searching for Top Squarks at the LHC in Fully Hadronic Final State}

\author{Bhaskar Dutta$^{1}$}
\author{Teruki Kamon$^{1,2}$}
\author{Nikolay Kolev$^{3}$}
\author{Kuver Sinha$^{1}$}
\author{Kechen Wang$^{1}$}

\affiliation{$^{1}$~Department of Physics, Texas A\&M University, College Station, TX 77843-4242, USA \\
$^{2}$~Department of Physics, Kyungpook National University, Daegu 702-701, South Korea \\
$^{3}$~Department of Physics, University of Regina, SK, S4S 0A2, Canada
}

\begin{abstract}

We pursue a scenario where the lighter top squark (stop) mass is accessible for the Large Hadron Collider (LHC) in the near future, while gluinos and first two generation squarks are too heavy. At $\sqrt{s} = 8$ TeV, we investigate the identification of stops which decay predominantly into a top quark and the stable lightest supersymmetric particle. We use a simple kinematical variable, $M3$, to reconstruct two top quarks which are pair-produced from the stops, in the fully hadronic channel.  The dominant Standard Model (SM) background for this signal stems from $t\bar t$ plus jets, with one top quark decaying into $ bl\nu$, where the lepton is undetected and the $\nu$ produces missing transverse momentum. The lepton identification efficiency is thus crucial in order to estimate the background correctly. We identify kinematical variables to reduce the SM background.
We find that it is possible to achieve signal and background cross-section at similar levels for stop masses around $350 - 500$ GeV for a $\tilde{\chi}^0_1$ mass of $100$ GeV.

\end{abstract}
MIFPA-12-25
\maketitle


\section{Introduction}

Experiments at the Large Hadron Collider (LHC) have recently reported preliminary evidence for a Higgs-like particle, with mass in the region of $125$ GeV \cite{ATLAS:2012ae, Chatrchyan:2012tx}.

In the Standard Model (SM), higher loop corrections to the Higgs mass are quadratically divergent; to avoid fine-tuning, new physics should appear around a scale of $\mathcal{O}(1)$ TeV and cut off the divergences. The problem is most severe in the case of the one-loop correction from the top sector, since other contributions to the Higgs mass are suppressed by gauge or smaller Yukawa couplings. Thus, reducing fine-tuning in the SM leads minimally to the conclusion that there should a partner for the top quark in the sub-TeV regime which is responsible for the cancellations. 

The most widely studied mechanism for canceling the divergences is supersymmetry, and in particular the dangerous top quark loops are canceled by the scalar superpartner of the top quark, called a stop ($\tilde{t}$).
 While the LHC has already started to put constraints on the first two generation squarks-gluino plane, the paucity of strong constraints on stop masses is due to the small production cross-section of stop pairs and a huge background from top quark production. The current constraints are: (a) for a directly produced stop going to top quark plus next to lightest neutralino, with neutralino decaying to gravitino plus $Z$, ATLAS excludes stop masses upto $240-330$ GeV using $2.05$ fb$^{-1}$ \cite{Aad:2012cz} (b) at the Tevatron, the stop mass constraint is about $150-180$ GeV \cite{Abazov:2008rc, TevatronConstraintCDF}. 

In this paper, we will probe a technique for stop searches in the following decay mode
\be \label{basicdecay1}
\tilde{t} \, \rightarrow \, t + \tilde{\chi}^0_1 \,\, ,
\ee
where the $\tilde{\chi}^0_1$ is the lightest neutralino, which we will take to be the lightest supersymmetric particle (LSP). In $R$-parity conserving models, the LSP is the main source of missing energy in the event. We will always be speaking about the lightest top quark superpartner $\tilde{t}_1$, which we will hereafter call $\tilde{t}$. We will not make any assumptions about the spectrum, except that the above decay mode is kinematically allowed and dominant.
%

The main challenge in such searches is the fact that the LHC is a top quark factory and distinguishing top quarks produced from stop decay, as opposed top quarks produced directly, can be very challenging. There are several established techniques of probing the $t\overline{t}$ system or identifying top quarks: 

$(i)$ \,\,$\alpha_T$, the Razor, and $M_{T2}$: These techniques rely on the identification of two hemispheres to maximize sensitivity in searches for a pair of heavy colored objects and their cascade decays. These are inclusive searches that do not rely on the reconstruction of top quark(s); they have larger signal acceptance, but larger background \cite{hemispheres1, hemispheres2, hemispheres3, hemispheres4, hemispheres5, hemispheres6}. 

$(ii)$\,\, Methods that reconstruct a top quark or two top quarks, followed by cuts on topology or kinematics to reduce the background. These methods have lower background, and lower acceptance of signal. 

We choose to explicitly reconstruct the two-top quark system. Within this class of searches, several options are available:

$(iia)$ \,\, $M3$: This is the invariant mass of trijet combinations with highest vectorially summed $p_T$. $M3$ has been used in top quark studies at CMS \cite{Chatrchyan:2011ew} and CDF \cite{CDFtopstudy};

$(iib)$ \,\, Kinematic Fits: One minimizes $\chi^2$ in dijet and trijet invariant masses, matching to $W$ and top quark masses; for example \cite{Blyweert:2012bq};

$(iic)$ \,\, Top Taggers: See \cite{Plehn:2011tg} for a recent review; 

$(iid)$ \,\,Bi-Event Subtraction Technique (BEST): This method has been used to identify top quark system in \cite{Dutta:2011gs}. Jet combinatorics is reduced by mixing events in hadronic decay chains.

For the two top quark system in Eq.~\ref{basicdecay1}, the highest $p_T$ jets are most likely to be from the top quark, if it is signal. Intuitively, $M3$ should work well in such a system (note that such an assumption is untrue in a SUSY environment with multiple superpartners undergoing decays, where the highest $p_T$ jets are probably not from the top quark; BEST works better there). 

In \cite{Plehn:2012pr, Kaplan:2012gd}, top taggers have been used to reconstruct top quarks coming from stop decay, while in \cite{Alves:2012ft}, shape analyses of $\met$-related distributions have been used to probe the stop system when the stop-LSP mass difference is degenerate with the top quark mass.


At the LHC, it is expected that the existence of stops will be indirectly established initially using inclusive jets + single lepton + $\met$ analysis, however once any excess is observed the direct evidence of the stop can be established though the existence of top quarks in the signal. Our goal in the paper is to establish the existence of two top quarks in the final states along with the missing energy in all hadronic channel. In order to make our analysis realistic we use \pgs \, detector simulation \cite{pgs} and consider  $W+ n$ jets, $Z+ n$ jets and $t \overline{t} + n$ jets, with $n \leq 6$  as well as single top $+$ jets backgrounds.  Interestingly we noticed that $t \overline{t} + (3-6)$ jets contribution to the background is comparable to $t \overline{t} + (1-2)$ jets. 

Our finding is that simple kinematical selections with the $M3$ variable is an effective tool for stop searches. At $\sqrt{s} = 8$ TeV, we achieve background and signal cross-sections at comparable levels for stop masses around $350 - 500$ GeV. 


The outline of this paper is as follows. In Section \ref{Strategy}, we outline our search strategy. In Section \ref{results}, we give our results in detail. The results are also summarized in Table \ref{CutSummary350100} appearing at the end of the paper. We end with our conclusions.

\section{Search Strategy} \label{Strategy}

We consider the fully hadronic mode
\be
pp \rightarrow \tilde{t} \tilde{t}^* \rightarrow (t\tilde{\chi}^0_1)(\overline{t}\tilde{\chi}^0_1) \rightarrow (bjj\tilde{\chi}^0_1)(\overline{b}jj\tilde{\chi}^0_1)
\ee

We investigate samples with at least four non $b$-tagged jets and at least two $b$-tagged jets, along with large missing energy. Signal events are generated with \isajet \, \cite{isajet} + \pythia \, \cite{pythia}, background events are generated with \alpgen \, \cite{Mangano:2002ea} + \pythia, followed by \pgs \, detector simulation.

The main source of missing energy for the $t\overline{t}$ background are neutrinos coming from the leptonic decay of a $W$, while for the signal the dominant source of missing energy is the neutralino. Clearly, after the missing energy cut, the most critical factor affecting the discrimination of signal over background in the fully hadronic mode is the lepton veto efficiency. Due to imperfections of the lepton veto, $t\overline{t}$ events with leptonic $W$ decay could be a dominant source of background.


\begin{figure}[!htp] 
\centering
\includegraphics[width=3.5in]{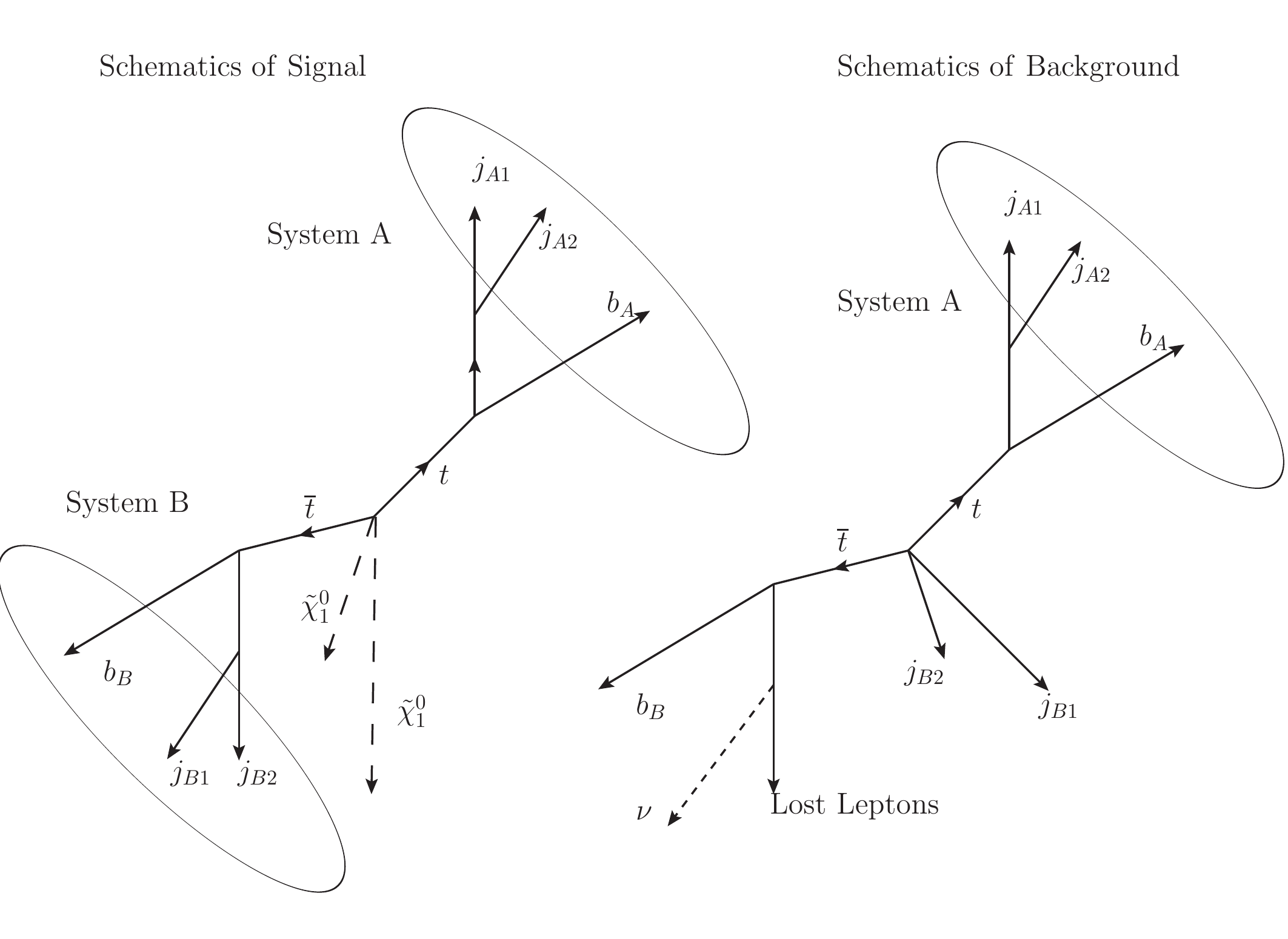}
\caption{[Left] schematic diagram of the signal. The stop pair gives rise to $t\overline{t}$ and neutralinos, which are the main source of $\met$. In the fully hadronic mode, the top quarks decay into trijet systems. ``System A" is the trijet system containing the leading $p_T$ jet and reconstructed using $M3$, while the remaining jets are called ``System B". [Right] $t\overline{t}$ background after lepton veto where the lepton is undetected. The main source of $\met$ here is the neutrino from $W$ decaying leptonically. The associated lepton passing the veto is termed a ``lost lepton".}
\label{StopSignalBG}
\end{figure}

Our method is to $(1)$ reconstruct a top quark using the trijet invariant mass $M3$, $(2)$ use kinematic correlations between the constituents of the two $(bjj)$ systems and $\met$ to improve the reconstruction of the pair of top quarks and $(3)$ finally apply $M3$ again to identify the second top quark. We describe these steps below, before showing our results in the next section.

$(1)$ \, We use $M3$ twice. First, combinations of three jets are made in the sample, keeping one $b$-tagged jet and two untagged jets in each trijet combination. Next, the trijet combination with the largest vectorially summed transverse momentum $p^{\rm leading}_{T,bjj}$ is chosen. 
%
%
%
%
The invariant mass of this trijet combination is defined as $M3(p^{\rm leading}_{T,bjj})$. It approximates the mass of the hadronically-decaying
top quark. Similarly, we find a $2^{\rm nd}$ leading trijet combination $p^{2^{\rm nd}}_{T,bjj}$ and invariant mass $M3(p^{2^{\rm nd}}_{T,bjj})$.
%
%
%
Associated with $M3$, we also define $M2$, which is the invariant mass of the two untagged jets in the trijet $M3$ combination. 

Using $M3$, we identify a first top quark, which we call ``System A". This is done by calculating $\chi^2$ for the trijet and dijet combination corresponding to the leading $p_T$ combination $M3(p^{\rm leading}_{T,bjj})$ and also for the $2^{\rm nd}$ leading combination $M3(p^{2^{\rm nd}}_{T,bjj})$, with a mean top quark mass of $170$ GeV and width of $15$ GeV, and a mean $W$ mass of $80$ GeV and a width of $10$  GeV. The combination with the lowest $\chi^2$ is then taken to represent System A. We call this combination $M3^{\rm min}$. 
%
%

We note that this $\chi^2$ analysis allows for more signal events in the identification of System A. We show the results of this analysis in Section \ref{M3cut}.

$(2)$ After the identification of System A, we classify the remaining $b$-jet and non-tagged jets to be ``System B"; thus, we would denote them as $(b_Bj_Bj_B)$. We employ various cuts on azimuthal angles between jets and $\met$, and $M_T$ between $b_B$ and $\met$. These are motivated by the fact that for signal, the main source of missing energy is the neutralino, while for the $t\overline{t}$ background, the main source of missing energy is the neutrino coming from the leptonic decay of the $W$, or from jet mismeasurement. Thus, for example, for the background, $\met$ is aligned along $b_B$, as is clear from the schematic diagram shown in Figure \ref{StopSignalBG}. For the stop decay, however, the correlation between the $\met$ in the form of neutralino and the $b_B$ is far weaker. The results of this analysis are shown in Section \ref{angularcuts}.

$(3)$ At the final stage, we employ $M3$ again to identify the second top quark, System B. The result of this analysis is shown in Section \ref{systemBcuts}.

\section{Results} \label{results}

In this section, we describe our selection criteria and the cross sections after every stage of cuts (see Table \ref{CutSummary350100}).

\subsection{Background}


We generate the following backgrounds with \alpgen + \pythia + \pgs: $t\overline{t}+ n$ jets with $n \leq 6$, single top $+$ jets, $W (rightarrow \tau \nu) + n$ jets and $Z (\rightarrow \nu \nu, \tau \tau) + n$ jets with $n$ upto $6$. The background cross-sections are given in Table \ref{Background1}. While Ref. \cite{Plehn:2012pr} considers  background with $t \overline{t} + (\leq) 2$ jets in the stop analysis, we find that  $t \overline{t} + (3-6)$ jets contribution is comparable to $t \overline{t} + (1-2)$ jets.



 

\begin{table}[!htp] 
\caption{Main sources of background. ``Others" includes single top $+$ jets, $W+n$ jets and $Z+n$ jets with $1 \leq n \leq 6$. All cross-sections are in fb.}
\label{Background1}
\begin{center}
\begin{tabular}{c| c c c  } \hline \hline

Background &$t\overline{t} + (\leq 2)j$ \,\, &$t\overline{t} + (3-6)j$ \,\, & Others \\ \\ 
Cross-section &$2.0 \times 10^5$  & $0.24 \times 10^5$  & $2.8 \times 10^6$         \\ 
 \hline \hline
\end{tabular}
\end{center}
\end{table}

%
%

\subsection{Baseline Cuts: $6$ jets, Lepton veto, $\met$}\label{baselinecuts}

Our baseline selection cuts are: 

$(i)$ $N_{{\rm non} \, b{\rm -jets}} \, \geq \, 4$, and at least two loosely tagged $b$-jets.

$(ii)$ The leading jet has $p_T \, > \, 100$ GeV in $|\eta| \leq 2.5$, and all other jets have $p_T \, > \, 30$ GeV in $|\eta| \leq 2.5$.

$(iii)$ Lepton veto: We reject isolated electrons and muons with $p_T > 10$ GeV in $|\eta| \leq 2.5$. The isolation criteria are $\sum p_{T \, {\rm iso}}^{\rm track} \, < \, 5$ GeV with $\Delta R = 0.4$.

$(iv)$ $\tau$ veto: We also reject any hadronically decaying $\tau$ with $p_T > 20$ GeV in $|\eta| \leq 2.1$. We assume a $\tau$ identification efficiency of $60\%$ and a fake rate of $2\%$.

$(v)$ $\met \, > \, 100$ GeV.

\subsection{$M3$: Tagging Top System A}\label{M3cut}

In this section, we use $M3$ to tag the top quark in System A, after a $\met$ cut to further reduce SM background. The value of the $\met$ cut is determined by maximizing the significance for each choice of mass. This is shown in Table \ref{metcutsformasses}.

\begin{table}[!htp] 
\caption{$\met$ cut for various choices of masses. All masses are in GeV.}
\label{metcutsformasses}
\begin{center}
\begin{tabular}{c| c c c c c | c c} \hline \hline

$\tilde{t}$             & $350$  & $400$  & $450$  &$500$  &$550$  &$400$  &$400$  \\ 
$\tilde{\chi}^0_1$      & $100$  & $100$  & $100$  &$100$  &$100$  &$150$  &$200$  \\  
\hline
$\met \,\, {\rm cut}$   & $145$  & $170$  & $195$ & $195$  & $195$  & $170$  & $100$      \\ 

 \hline \hline
\end{tabular}
\end{center}
\end{table} 

\begin{figure}[!htp] 
\centering
\includegraphics[width=3.5in]{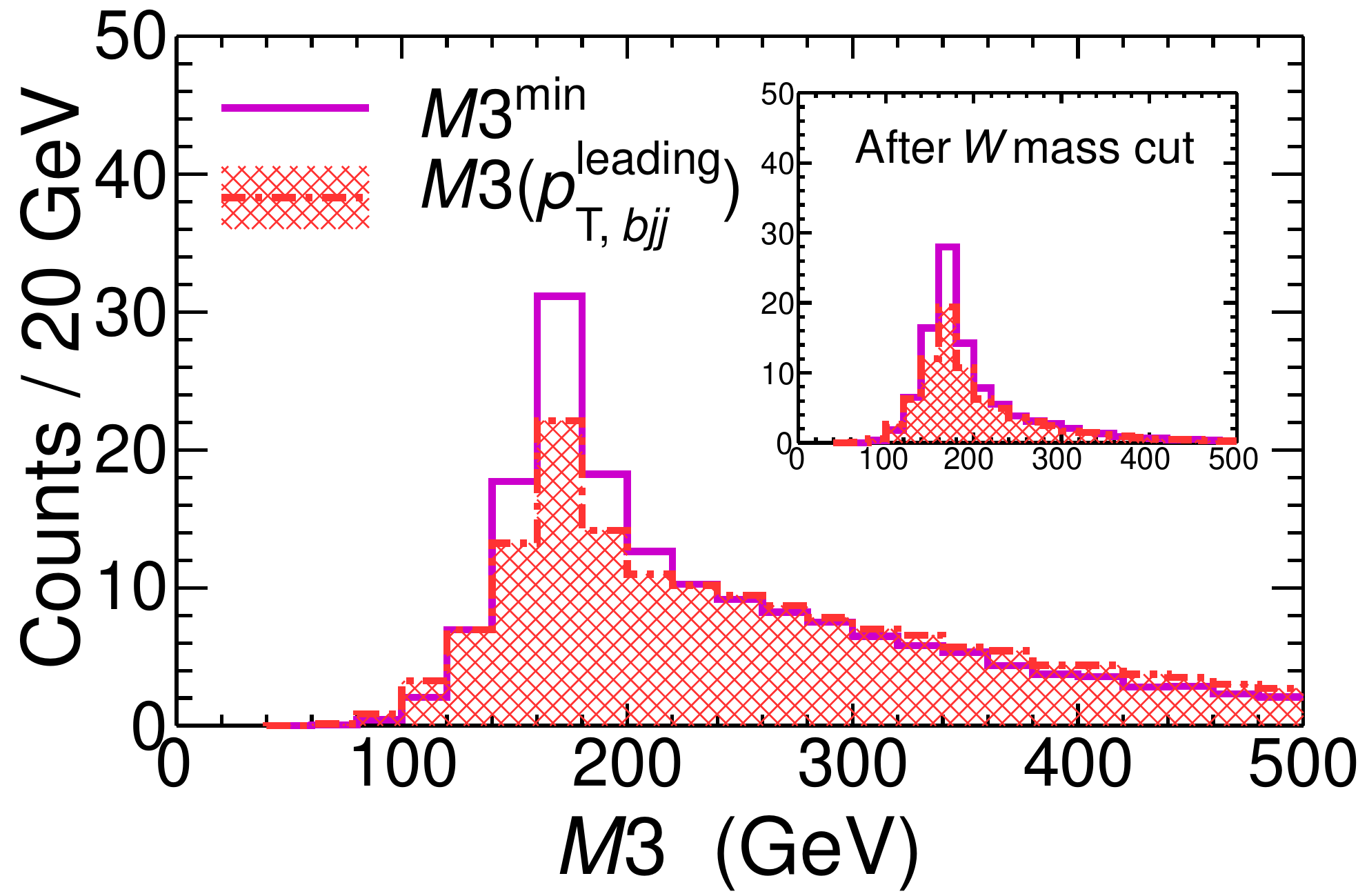}
\caption{Distributions of $M3^{\rm min}$ and $M3(p^{\rm leading}_{T,bjj})$. The inset shows the distribution after $M2$ mass window cut. A gain of $\sim 30\%$ in signal is obtained by using $M3^{\rm min}$. The luminosity is $50$ fb$^{-1}$.}
\label{M3chisquare}
\end{figure}

As described in Section \ref{Strategy}, we identify System A by using $M3$.  Figure \ref{M3chisquare} shows the comparative distributions of $M3^{\rm min}$ and $M3(p^{\rm leading}_{T,bjj})$. We improve the top tagging by approximately $30 \%$ in signal events in the top quark mass region.


%
%

We next perform a $W$ mass window cut on $M2^{\rm min}$, taking $40 \, {\rm GeV} \, \leq \, M2^{\rm min} \leq \, 120 \, {\rm GeV}$ and a top quark mass window cut on $M3^{\rm min}$, taking $120 \, {\rm GeV} \, \leq \, M3^{\rm min} \leq \, 220 \, {\rm GeV}$.
%
%
%
%

We show the $M3^{\rm min}$ distribution after $M2^{\rm min}$ mass cut in the inset of Figure \ref{M3chisquare}.


We now proceed to probe the constituents of the ``other top quark" in System B.

\subsection{Angular and $M_T$ Cuts: Kinematic Correlations between $\met$ and Jets}\label{angularcuts}

\begin{figure}[!htp] 
\centering
\includegraphics[width=3.5in]{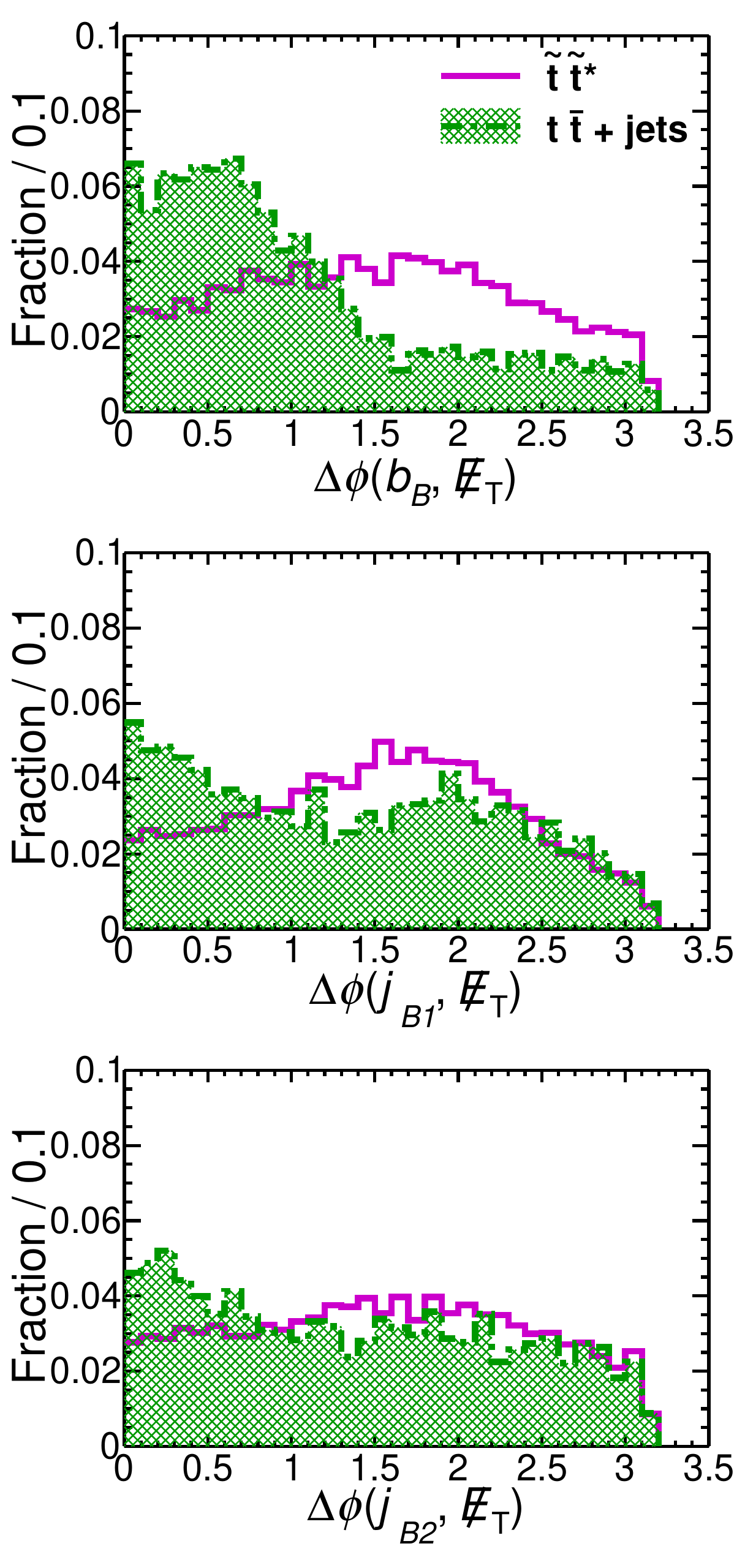}
\caption{Distributions of $\Delta \phi (b_B, \met), \Delta \phi (j_{B1}, \met), \Delta \phi (j_{B2}, \met)$ for  $t\overline{t}$ background and signal $(m_{\tilde{t}} = 400 \, {\rm GeV}, m_{\tilde{\chi}^0_1} = 100 \, {\rm GeV})$. We cut at $\Delta \phi (b_B, \met) \, > \, 1.2, \Delta \phi (j_{B(1,2)}, \met) \, > \, 0.7$. Here, $b_B, j_{B1},$ and $j_{B2}$ denote the $b$, leading jet, and next leading jet of System B. The luminosity is $50$ fb$^{-1}$.}
\label{dPhiMETbB}
\end{figure}

\begin{figure}[!htp] 
\centering
\includegraphics[width=3.5in]{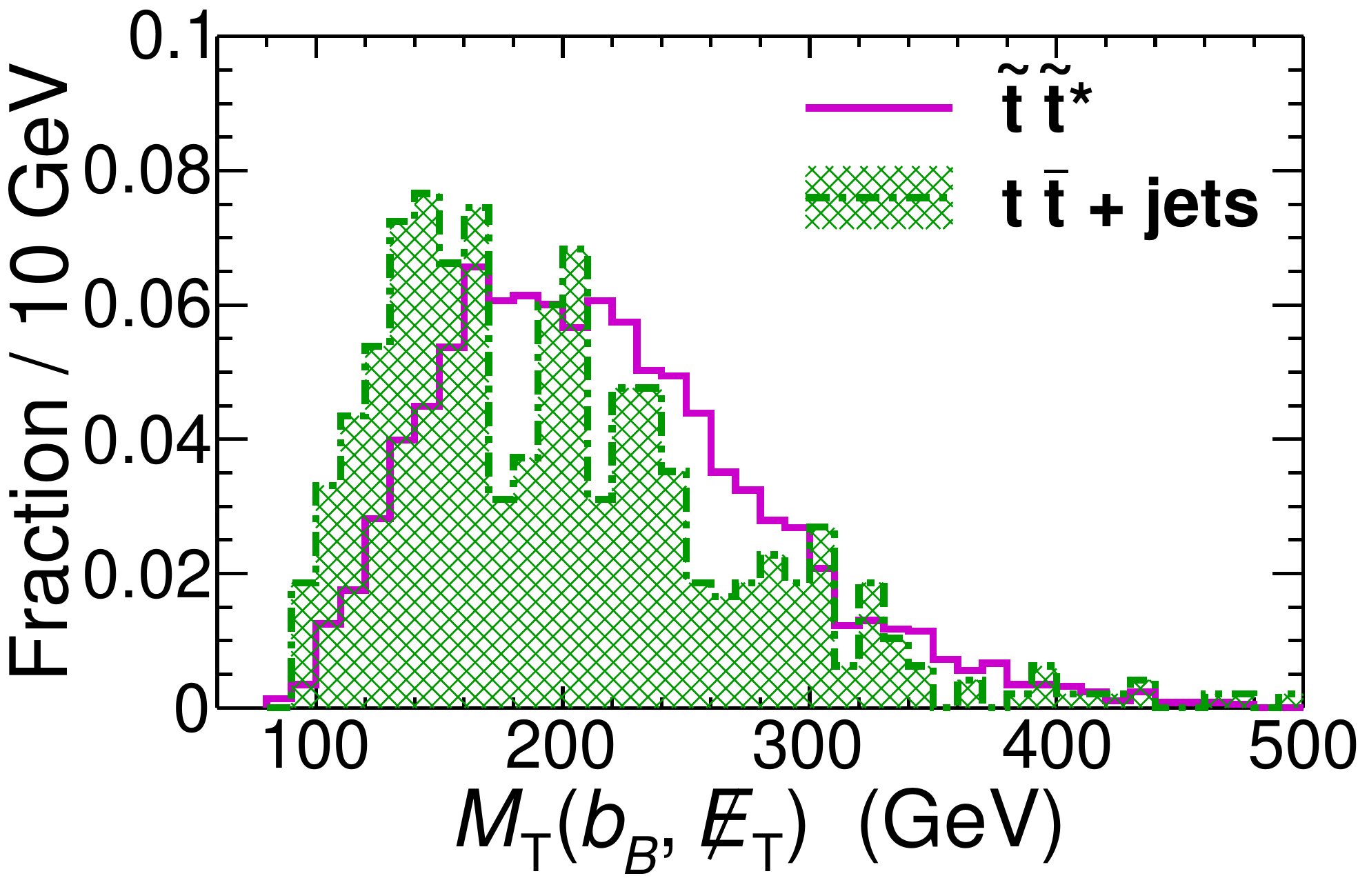}
\caption{Distribution of $M_T(b_B, \met)$ for  $t\overline{t}$ background and signal $(m_{\tilde{t}} = 400 \, {\rm GeV}, m_{\tilde{\chi}^0_1} = 100 \, {\rm GeV})$. We cut at $M_T(b_B, \met) \, > \, 155$ \, GeV. The luminosity is $50$ fb$^{-1}$.}
\label{MTMETbB}
\end{figure}

We denote the remaining $b$-jet and non-tagged jets as $(b_Bj_Bj_B)$. We clean up the system with various angular and $M_T$ cuts, as mentioned in our search strategy in Section \ref{Strategy}. The cuts values arechosen based on Figures \ref{dPhiMETbB} and \ref{MTMETbB}:

$(i)\,\,$ $\Delta \phi (b_B, \met) \, > \, 1.2$ and $\Delta \phi (j_{B(1,2)}, \met) \, > \, 0.7$, where $j_{B(1,2)}$ refer to the first and second leading jets in System B, respectively. 



$(ii)$ \,\, $M_T(b_B, \met):$  We choose optimal cut values for different masses (see Table \ref{MTbprimecutsformasses}).

\begin{table}[!htp] 
\caption{$M_T(b_B, \met)$ cuts for various choices of masses. All masses are in GeV.}
\label{MTbprimecutsformasses}
\begin{center}
\begin{tabular}{c| c c c c c | c c} \hline \hline

$\tilde{t}$             & $350$  & $400$  & $450$  &$500$  &$550$  &$400$  &$400$  \\ 
$\tilde{\chi}^0_1$      & $100$  & $100$  & $100$  &$100$  &$100$  &$150$  &$200$  \\  
\hline
$M_T(b_B, \met)$         & $145$  & $155$  & $165$  &$165$  &$165$  &$155$  &$155$  \\ 

 \hline \hline
\end{tabular}
\end{center}
\end{table}

%

\begin{figure}[!htp] 
\centering
\includegraphics[width=3.5in]{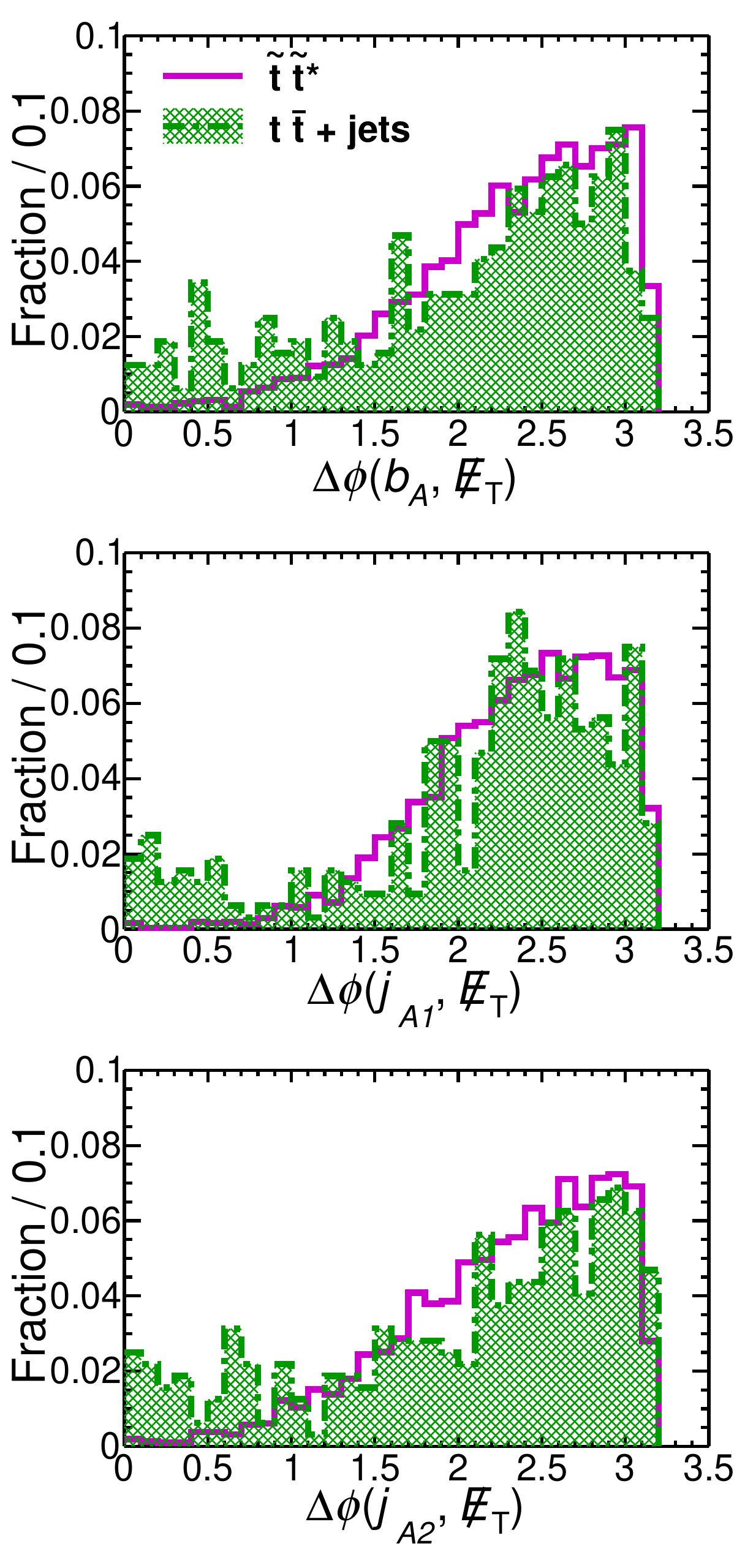}
\caption{Distributions of $\Delta \phi (b_A, \met)$, $\Delta \phi (j_{A1}, \met)$, and $\Delta \phi (j_{A2}, \met)$ for $t\overline{t}$ background and signal $(m_{\tilde{t}} = 400 \, {\rm GeV}, m_{\tilde{\chi}^0_1} = 100 \, {\rm GeV})$. We cut at $\Delta \phi (b_A, \met) \, > \, 1.2$ and $\Delta \phi (j_{A(1,2)}, \met) \, > \, 0.7$. Here, $b_A, j_{A1},$ and $j_{A2}$ denote the $b$, leading jet, and next leading jet of System A. The luminosity is $50$ fb$^{-1}$.}
\label{dPhiMETbA}
\end{figure}

After the above cuts, we revert to the trijet $(b_Aj_Aj_A)$ in System A with similar angular cuts between missing energy and the $b$-tagged jet as well as non-tagged jets. These angular cuts are efficient in reducing events with ``lost leptons". The cuts are chosen based on Figure \ref{dPhiMETbA}:

$(iii)$ \,\, $\Delta \phi (b_A, \met) \, > \, 1.2$ and $\Delta \phi (j_{A(1,2)}, \met) \, > \, 0.7$, where $j_{A(1,2)}$ refer to the first and second leading jets in System A, respectively. 



%

\subsection{$M3:$ Tagging Top System B}\label{systemBcuts}

\begin{figure}[!htp] 
\centering
\includegraphics[width=3.5in]{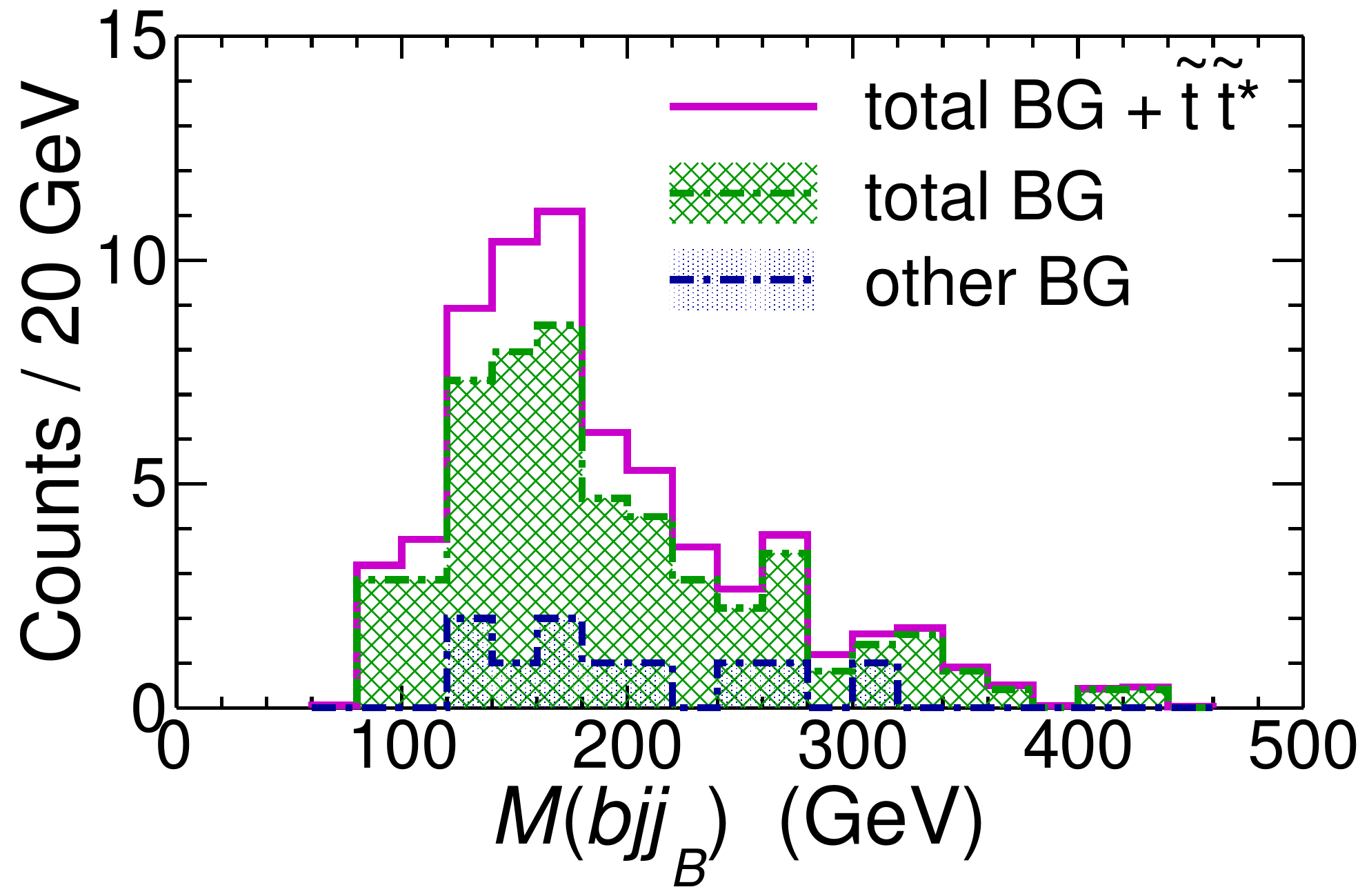}
\caption{Distribution of $M3$ in System B, after requiring $40 \, {\rm GeV} \, \leq \, M2 \leq \, 120 \, {\rm GeV}$. Displayed are: ``other" sources of background (single top $+$ jets, $W+n$ jets and $Z+n$ jets with $n \leq 6$), total background including $t\overline{t}+n$ jets, and total background plus signal for our reference point $(m_{\tilde{t}} = 400 \, {\rm GeV}, m_{\tilde{\chi}^0_1} = 100 \, {\rm GeV})$. The luminosity is $50$ fb$^{-1}$.}
\label{TopsystemB}
\end{figure}

As a last step, $M3$ is applied in System B, followed by a $W$ mass window cut on $M2$ ($40 \, {\rm GeV} \, \leq \, M2 \leq \, 120 \, {\rm GeV}$). The $M3$ distribution is shown in Figure \ref{TopsystemB}. Our final results with $110 \, {\rm GeV} \, \leq \, M3 \leq \, 230 \, {\rm GeV}$ are tabulated in Table \ref{CutSummary350100}.

We note that for the point $(m_{\tilde{t}} = 350 \, {\rm GeV}, m_{\tilde{\chi}^0_1} = 100 \, {\rm GeV})$ we additionally impose $\Delta \phi (b_{A,B}, \met) \, < \, 2.7$. Also, for the point $(m_{\tilde{t}} = 400 \, {\rm GeV}, m_{\tilde{\chi}^0_1} = 200 \, {\rm GeV})$, the $W$ mass window cut on $M2$ of System B was taken as $60 \, {\rm GeV} \, \leq \, M2 \leq \, 100 \, {\rm GeV}$, while the top quark mass window was taken as $140 \, {\rm GeV} \, \leq \, M3 \leq \, 200 \, {\rm GeV}$.


%
%
%
%

\section{Summary and Conclusion}


In this paper, we have explored a search strategy for a light stop at the LHC at $8$ TeV, using $M3$ variable in the fully hadronic channel. The gluino and the first two generation squarks are assumed to be too heavy to be produced significantly at the LHC. 

We first performed $M3^{\rm min}$ to identify a top quark system (System A). Next, we performed $M3$ again to identify the second top quark (System B), along with a series of kinematical cuts to reduce the SM backgrounds. Throughout this study, we used \pgs \, detector simulation and considered  $W+ n$ jets, $Z+ n$ jets and $t \overline{t} + n$ jets, with $n \leq 6$  as well as single top $+$  jets backgrounds.  Interestingly we noticed that $t \overline{t} + (3-6)$ jets contribution to the background is comparable to $t \overline{t} + (1-2)$ jets.


Tables \ref{SummaryTable} and \ref{CutSummary350100} are a summary of the search performance for various choices of stop and neutralino masses. 

\begin{table}[!htp] 
\caption{Final significances for various choices of masses. All masses are in GeV. The luminosity is $50$ fb$^{-1}$.}
\label{SummaryTable}
\begin{center}
\begin{tabular}{c| c c c c c | c c } \hline \hline

$\tilde{t}$             & $350$    & $400$   & $450$   &$500$   &$550$   &$400$   &$400$ \\ 
$\tilde{\chi}^0_1$      & $100$    & $100$   & $100$   &$100$   &$100$   &$150$   &$200$\\  
\hline
$S/\sqrt{B}$            & $1.29$   & $1.71$  & $1.39$  &$0.81$  &$0.35$  &$0.94$  &$0.47$      \\ 

 \hline \hline
\end{tabular}
\end{center}
\end{table} 



In summary, a simple kinematical reconstruction technique using the $M3$ variable is an effective tool for stop searches. We find that at $\sqrt{s} = 8$ TeV it is possible to reduce background down to a level of signal cross-section for stop masses around $350 - 500$ GeV for a $\tilde{\chi}^0_1$ mass of $100$ GeV. 

\section{Acknowledgements}

We would like to thank Matt Buckley, Christopher Lester, Michelangelo Mangano, Tilman Plehn, Michael Spannowsky,  Daniel Stolarski and Michihisa Takeuchi for very useful discussions and correspondence, and Tai Sakuma for collaboration in the early stages of this work. We would like to thank John Conway for help with \pgs \,. We would also like to thank David Curtin and Will Flanagan for correspondence. This work is supported in part by the DOE grant DE-FG02-95ER40917 and by the World Class University (WCU) project through the National Research Foundation (NRF) of Korea funded by the Ministry of Education, Science \& Technology (grant No. R32-2008-000-20001-0).

\onecolumngrid

\begin{table}[!htp] 
\caption{Summary of effective cross sections (fb) for stop pair production and the SM background events in our stop search feasibility study. Masses and momenta are in GeV. ``Other" sources of background include  single top $+$ jets, $W+n$ jets and $Z+n$ jets with $1 \leq n \leq 6$. The significance is given at $50$ fb $^{-1}$.}
\label{CutSummary350100}
\begin{center}
\begin{tabular}{c |c c c c c} 
\hline \hline 
                    

          &                      &Signal    &\,\, $t\overline{t}+ n (\leq 2)$jets &\,\, $t\overline{t}+ n(\geq 3)$jets &\,\, Others \\
          
\hline 
              \hline 
                    


                     &Initial                             & $760$ &$2.0 \times 10^5$ &$0.24 \times 10^5$ &$2.8 \times 10^6$     \\

      &Baseline Cuts (Sec. \ref{baselinecuts})            &$9.96$   &\,\, $192$                        &\,\,$147$                      &\,\,$31.7$      \\      

$m_{\tilde{t}} = 350$   &$\met \, > \, 145$ GeV                      &$6.79$   &\,\, $82.2$                        &\,\,$69.1$                      &\,\,$15.8$      \\

 $m_{\tilde{\chi}^0_1} = 100$                &System A: $M3$ (Sec. \ref{M3cut})                                                 &$2.65$  &\,\,$25.8$                          &\,\,$15.6$                      &\,\,$3.14$  \\ 

            & Angular and $M_T$ cuts (Sec. \ref{angularcuts})                                                 &$0.55$  &\,\,$1.61$                          &\,\,$1.71$                      &\,\,$0.72$   \\

              & System B: $M3$ (Sec. \ref{systemBcuts})                                                 &$0.25$  &\,\,$0.40$                          &\,\,$0.47$                      &\,\,$0.20$                    \\


& $\Delta \phi (b_{A,B}, \met) \, < \, 2.7$            &$0.14$  &\,\,$0.24$                          &\,\,$0.25$                      &\,\,$0.10$   \\

  & Significance $(S/\sqrt{B})$                       & \multicolumn{4}{c}{$1.29$} \\

\hline  
                    


                     &Initial                             & $337$  &$2.0 \times 10^5$ &$0.24 \times 10^5$ &$2.8 \times 10^6$     \\

      &Baseline Cuts (Sec. \ref{baselinecuts})            &$5.55$   &\,\, $192$                        &\,\,$147$                      &\,\,$31.7$      \\      

$m_{\tilde{t}} = 400$   &$\met \, > \, 170$ GeV                      &$3.62$   &\,\, $53,4$                        &\,\,$47.0$                      &\,\,$11.1$      \\

$m_{\tilde{\chi}^0_1} = 100$                 &System A: $M3$ (Sec. \ref{M3cut})                                                 &$1.46$  &\,\,$15.4$                          &\,\,$9.73$                      &\,\,$1.82$  \\ 

            & Angular and $M_T$ cuts (Sec. \ref{angularcuts})                                                 &$0.44$  &\,\,$0.96$                          &\,\,$1.06$                      &\,\,$0.54$   \\

              &System B: $M3$ (Sec. \ref{systemBcuts})                                                 &$0.20$  &\,\,$0.26$                          &\,\,$0.28$                      &\,\,$0.14$                    \\


  & Significance $(S/\sqrt{B})$                       & \multicolumn{4}{c}{$1.71$} \\


\hline 
                    


                     &Initial                             & $160$  &$2.0 \times 10^5$ &$0.24 \times 10^5$ &$2.8 \times 10^6$     \\

      &Baseline Cuts (Sec. \ref{baselinecuts})            &$2.52$   &\,\, $192$                        &\,\,$147$                      &\,\,$31.7$      \\      

$m_{\tilde{t}} = 450$   &$\met \, > \, 195$ GeV                      &$1.61$   &\,\, $34.5$                        &\,\,$31.9$                      &\,\,$8.08$      \\

$m_{\tilde{\chi}^0_1} = 100$                 &System A: $M3$ (Sec. \ref{M3cut})                                                 &$0.62$  &\,\,$9.17$                          &\,\,$6.32$                      &\,\,$1.30$  \\ 

            & Angular and $M_T$ cuts (Sec. \ref{angularcuts})                                                 &$0.25$  &\,\,$0.55$                          &\,\,$0.69$                      &\,\,$0.40$   \\

              &System B: $M3$ (Sec. \ref{systemBcuts})                                                 &$0.12$  &\,\,$0.17$                          &\,\,$0.14$                      &\,\,$0.06$                    \\


  & Significance $(S/\sqrt{B})$                       & \multicolumn{4}{c}{$1.39$} \\
%
 \hline 
                    


                     &Initial                             & $80.5$  &$2.0 \times 10^5$ &$0.24 \times 10^5$ &$2.8 \times 10^6$     \\

      &Baseline Cuts (Sec. \ref{baselinecuts})            &$1.21$   &\,\, $192$                        &\,\,$147$                      &\,\,$31.7$      \\      

$m_{\tilde{t}} = 500$   &$\met \, > \, 195$ GeV                      &$0.86$   &\,\, $34.5$                        &\,\,$31.9$                      &\,\,$8.08$      \\

$m_{\tilde{\chi}^0_1} = 100$                 &System A $M3$ (Sec. \ref{M3cut})                                                 &$0.32$  &\,\,$9.17$                          &\,\,$6.32$                      &\,\,$1.30$  \\ 

            & Angular and $M_T$ cuts (Sec. \ref{angularcuts})                                                 &$0.15$  &\,\,$0.55$                          &\,\,$0.69$                      &\,\,$0.40$  \\ 

              &System B: $M3$ (Sec. \ref{systemBcuts})                                                 &$0.07$  &\,\,$0.17$                          &\,\,$0.14$                      &\,\,$0.06$                   \\


  & Significance $(S/\sqrt{B})$                       & \multicolumn{4}{c}{$0.81$} \\


\hline 

                     &Initial                             & $43.0$  &$2.0 \times 10^5$ &$0.24 \times 10^5$ &$2.8 \times 10^6$     \\

      &Baseline Cuts (Sec. \ref{baselinecuts})            &$0.57$   &\,\, $192$                        &\,\,$147$                      &\,\,$31.7$      \\      

$m_{\tilde{t}} = 550$   &$\met \, > \, 195$ GeV                      &$0.43$   &\,\, $34.5$                        &\,\,$31.9$                      &\,\,$8.08$      \\

$m_{\tilde{\chi}^0_1} = 100$                 &System A: $M3$ (Sec. \ref{M3cut})                                                 &$0.14$  &\,\,$9.17$                          &\,\,$6.32$                      &\,\,$1.30$  \\ 

            & Angular and $M_T$ cuts (Sec. \ref{angularcuts})                                                 &$0.07$  &\,\,$0.55$                          &\,\,$0.69$                      &\,\,$0.40$  \\ 

              &System B: $M3$ (Sec. \ref{systemBcuts})                                                 &$0.03$  &\,\,$0.17$                          &\,\,$0.14$                      &\,\,$0.06$  \\                  


  & Significance $(S/\sqrt{B})$                       & \multicolumn{4}{c}{$0.35$} \\

              \hline \hline 
                    


                     &Initial                             & $337$  &$2.0 \times 10^5$ &$0.24 \times 10^5$ &$2.8 \times 10^6$     \\

      &Baseline Cuts (Sec. \ref{baselinecuts})            &$4.78$   &\,\, $192$                        &\,\,$147$                      &\,\,$31.7$      \\      

$m_{\tilde{t}} = 400$   &$\met \, > \, 170$ GeV                      &$2.76$   &\,\, $53.4$                        &\,\,$47.0$                      &\,\,$11.1$      \\

$m_{\tilde{\chi}^0_1} = 150$                 &System A: $M3$ (Sec. \ref{M3cut})                                                 &$1.01$  &\,\,$15.4$                          &\,\,$9.73$                      &\,\,$1.82$  \\ 

            & Angular and $M_T$ cuts (Sec. \ref{angularcuts})                                                 &$0.23$  &\,\,$0.96$                          &\,\,$1.06$                      &\,\,$0.54$   \\

              & System B: $M3$ (Sec. \ref{systemBcuts})                                                 &$0.11$  &\,\,$0.26$                          &\,\,$0.28$                      &\,\,$0.14$                    \\

  & Significance $(S/\sqrt{B})$                       & \multicolumn{4}{c}{$0.94$} \\ 
  
              \hline 
                    


                     &Initial                             & $337$  &$2.0 \times 10^5$ &$0.24 \times 10^5$ &$2.8 \times 10^6$     \\

$m_{\tilde{t}} = 400$      &Baseline Cuts (Sec. \ref{baselinecuts})            &$3.34$   &\,\, $192$                        &\,\,$147$                      &\,\,$31.7$      \\      


$m_{\tilde{\chi}^0_1} = 200$                 &System A: $M3$ (Sec. \ref{M3cut})                                                 &$1.13$  &\,\,$67.2$                          &\,\,$38.8$                      &\,\,$7.40$  \\ 

            & Angular and $M_T$ cuts (Sec. \ref{angularcuts})                                                 &$0.87$  &\,\,$45.8$                          &\,\,$28.3$                      &\,\,$6.04$   \\

              & System B: $M3$ (Sec. \ref{systemBcuts})                                                 &$0.18$  &\,\,$4.12$                          &\,\,$2.59$                      &\,\,$0.54$                    \\

  & Significance $(S/\sqrt{B})$                       & \multicolumn{4}{c}{$0.47$} \\

\hline \hline

\end{tabular}
\end{center}
\end{table}

\twocolumngrid



\begin{thebibliography}{17}

\bibitem{ATLAS:2012ae} 
  G.~Aad {\it et al.}  [ATLAS Collaboration],
  ``Combined search for the Standard Model Higgs boson using up to 4.9 fb$^{-1}$ of $pp$ collision data at $\sqrt{s} = 7$ TeV with the ATLAS detector at the LHC,''
  Phys.\ Lett.\ B {\bf 710}, 49 (2012)
  [arXiv:1202.1408 [hep-ex]].


\bibitem{Chatrchyan:2012tx} 
  S.~Chatrchyan {\it et al.}  [CMS Collaboration],
  ``Combined results of searches for the standard model Higgs boson in $pp$ collisions at $\sqrt{s} = 7$ TeV,''
  Phys.\ Lett.\ B {\bf 710}, 26 (2012)
  [arXiv:1202.1488 [hep-ex]].



\bibitem{Aad:2012cz} 
  G.~Aad {\it et al.}  [ATLAS Collaboration],
  ``Search for scalar top quark pair production in natural gauge mediated supersymmetry models with the ATLAS detector in $pp$ collisions at $\sqrt{s} = 7$ TeV,''
  arXiv:1204.6736 [hep-ex].

  
\bibitem{Abazov:2008rc} 
  V.~M.~Abazov {\it et al.}  [D0 Collaboration],
  ``Search for scalar top quarks in the acoplanar charm jets and missing transverse energy final state in $p \bar{p}$ collisions at $\sqrt{s} = 1.96$-TeV,''
  Phys.\ Lett.\ B {\bf 665}, 1 (2008)
  [arXiv:0803.2263 [hep-ex]].
  
\bibitem{TevatronConstraintCDF}
  [CDF collaboration], 
  CDF Public Note 9834.
  

\bibitem{hemispheres1} 
  [CMS Collaboration],
  ``Search for supersymmetry in all-hadronic events with $\alpha_T$,''
  CMS-PAS-SUS-11-003.  
  
  \bibitem{hemispheres2} 
  L.~Randall and D.~Tucker-Smith,
  ``Dijet Searches for Supersymmetry at the LHC,''
  Phys.\ Rev.\ Lett.\  {\bf 101}, 221803 (2008)
  [arXiv:0806.1049 [hep-ph]].  

\bibitem{hemispheres3} 
    S.~Chatrchyan {\it et al.}  [CMS Collaboration],
  ``Inclusive search for squarks and gluinos in $pp$ collisions at $\sqrt{s} = 7$ TeV,''
  Phys.\ Rev.\ D {\bf 85}, 012004 (2012)
  [arXiv:1107.1279 [hep-ex]]. 
  
\bibitem{hemispheres4}   
  C.~Rogan,
  ``Kinematical variables towards new dynamics at the LHC,''
  arXiv:1006.2727 [hep-ph]. 
  
  \bibitem{hemispheres5} 
  T.~Aaltonen {\it et al.}  [CDF Collaboration],
  ``Top Quark Mass Measurement using mT2 in the Dilepton Channel at CDF,''
  Phys.\ Rev.\ D {\bf 81}, 031102 (2010)
  [arXiv:0911.2956 [hep-ex]]. 
  
  \bibitem{hemispheres6} 
   H.~Weber {\it et al.}  [CMS Collaboration],
  ``Search for supersymmetry in hadronic final states with MT2,''
  arXiv:1201.4659 [hep-ex].
  
  
\bibitem{Chatrchyan:2011ew} 
  S.~Chatrchyan {\it et al.}  [CMS Collaboration],
  ``Measurement of the Top-antitop Production Cross Section in $pp$ Collisions at $\sqrt{s}=7$ TeV using the Kinematic Properties of Events with Leptons and Jets,''
  Eur.\ Phys.\ J.\ C {\bf 71}, 1721 (2011)
  [arXiv:1106.0902 [hep-ex]].

\bibitem{CDFtopstudy} 
CDF Collaboration, 
``A Measurement of Top Quark Mass Using MET + Jets Events with $5.7$ fb$^{-1}$ of CDF Data,"
CDF public note 10433 (2011).
  Z.~Ye [CDF and D0 Collaborations],
  ``Top Quark Mass Measurements at the Tevatron,''
  arXiv:1107.4539 [hep-ex].

\bibitem{Dutta:2011gs} 
  B.~Dutta, T.~Kamon, N.~Kolev and A.~Krislock,
  ``Bi-Event Subtraction Technique at Hadron Colliders,''
  Phys.\ Lett.\ B {\bf 703}, 475 (2011)
  [arXiv:1104.2508 [hep-ph]].
  
\bibitem{Blyweert:2012bq} 
  S.~Blyweert [ATLAS and CMS Collaborations],
  ``Top-quark mass measurements at the LHC,''
  arXiv:1205.2175 [hep-ex].
  
\bibitem{Plehn:2011tg} 
  T.~Plehn and M.~Spannowsky,
  ``Top Tagging,''
  arXiv:1112.4441 [hep-ph].

\bibitem{Plehn:2012pr} 
  T.~Plehn, M.~Spannowsky and M.~Takeuchi,
  ``Stop searches in 2012,''
  arXiv:1205.2696 [hep-ph].

\bibitem{Kaplan:2012gd} 
  D.~E.~Kaplan, K.~Rehermann and D.~Stolarski,
  ``Searching for Direct Stop Production in Hadronic Top Data at the LHC,''
  arXiv:1205.5816 [hep-ph].

\bibitem{Alves:2012ft} 
  D.~S.~M.~Alves, M.~R.~Buckley, P.~J.~Fox, J.~D.~Lykken and C.~-T.~Yu,
  ``Stops and MET: the shape of things to come,''
  arXiv:1205.5805 [hep-ph].

\bibitem{pgs}
\pgs\ is a parameterized detector simulator.
We use version 4
(\url{http://www.physics.ucdavis.edu/~conway/research/software/pgs/pgs4-general.htm})
in the CMS detector configuration.

\bibitem{isajet}
  F. E. Paige, S.~D.~Protopopescu, H.~Baer and X.~Tata,
  ``ISAJET 7.69: A Monte Carlo event generator for p p, anti-p p, and e+ e- reactions,''
  [hep-ph/0312045].
 We use \isajet version 7.74.

\bibitem{pythia}
T. Sjostrand, S. Mrenna, and P. Skands,
 "PYTHIA 6.4 Physics and Manual."
  J. High Energy Phys. \textbf{05} (2006) 026.
  We use \pythia\ version 6.411 with TAUOLA.

\bibitem{Mangano:2002ea} 
  M.~L.~Mangano, M.~Moretti, F.~Piccinini, R.~Pittau and A.~D.~Polosa,
  ``ALPGEN, a generator for hard multiparton processes in hadronic collisions,''
  JHEP {\bf 0307}, 001 (2003)
  [hep-ph/0206293].


\end{thebibliography}
\end{document}